# Geometry-influenced cooling performance of lithium-ion battery

**Dwijendra Dubey[a], A. Mishra[a*], Subrata Ghosh[b], M.V. Reddy[c], Ramesh Pandey[a]**

[a]*Department of Applied Mechanics, Motilal Nehru National Institute of Technology Allahabad, Prayagraj, 211004, Uttar Pradesh, India*
[b]*Micro and Nanostructured Materials Laboratory – Nano Lab, Department of Energy, Politecnico di Milano, via Ponzio 34/3, Milano-20133, Italy*
[c]*Nouveau Monde Graphite (NMG), Saint-Michel-des- Saints, Québec, J0K 3B0, Canada*

*Corresponding Author: ashutoshjssate@gmail.com, amishra@mnnit.ac.in*

## Abstract

Battery geometry (shape and size) is one of the important parameters which governs the battery capacity and thermal behavior. In the dynamic conditions or during the operation, the performance of batteries become much more complex. Herein, the changes in thermal behavior of lithium-ion battery (LIB)by altering the geometry i.e., length to diameter ratio (*l/d*), is investigated. The geometries considered are named as large geometry (LG), datum geometry (DG) and small geometry (SG) with the *l/d* ratio of 5.25, 3.61, and 2.38, respectively. A three-dimensional (3D) multi-partition thermal model is adopted, and the numerical results are validated by the published experimental data. For three different cooling approaches such as radial, both-tab and mixed cooling, the average battery temperature and temperature heterogeneity are thoroughly examined considering the heat transfer coefficients (h) of50 and 100 W/m$^2$K at discharge rates of 1, 2 and 3C. Amongst, the minimum average battery temperature is exhibited by DG, the minimum radial temperature heterogeneity is obtained from LG, and substantial outperformance in terms of faster cooling rate is identified for SG, irrespective of the cooling approach employed.

Nomenclature

| Symbols | Descriptions | Symbols | Descriptions |
|---|---|---|---|
| $DOD$ | Depth of discharge | $T_o$ | Ambient temperature |
| $SOC$ | State of charge | $R$ | Radius |
| $LIB$ | Lithium-ion battery | $r_o$ | Maximum radius |
| $DG$ | Datum geometry | $H$ | Convective heat transfer coefficient |
| $LG$ | Large geometry | $V_{cell}$ | Operating potential of battery |
| $SG$ | Small geometry | $I$ | Current |
| $3D$ | Three dimensional | $U$ | Equilibrium potential |
| $(l/d)$ ratio | Length to diameter ratio | $Q$ | Heat |
| $OCP$ | Open circuit potential | $Q_{gen}$ | Internal heat generation |
| $AST$ | Average surface temperature | $Q_{jr}$ | Heat generation of jelly-roll |
| $RTD$ | Radial temperature difference | $Q_{pos}$ | Heat generation of positive terminal |
| $ATD$ | Axial temperature difference | $Q_{vol}$ | Volumetric heat generation |
| $T$ | Time | $K$ | Thermal conductivity |
| $C_p$ | Heat capacity at constant pressure | $K_r$ | Radial thermal conductivity |
| $P$ | Density | $K_q$ | Polar thermal conductivity |
| $R_{pos}$ | Resistance of positive terminal | $K_z$ | Axial thermal conductivity |
| $R_{neg}$ | Resistance of negative terminal | $(K_z)_{equivalent}$ | Equivalent axial thermal conductivity |
| $R_{cap}$ | Resistance of cap | $(K_r)_{equivalent}$ | Equivalent radial thermal conductivity |
| $R_{jr\ o}$ | Ohmic resistance of jelly-roll | $(R_{thermal})_{radial}$ | Radial thermal resistance |
| $R_{jr\ p}$ | Polarization resistance of jelly-roll | $(R_{thermal})_{axial}$ | Axial thermal resistance |
| $T_{in}$ | Temperature inside of LIB | $K_{r1}$ | Radial thermal conductivity of negative terminal |
| $T_{out}$ | Temperature at outer surface of battery | $K_{r2}$ | Radial thermal conductivity of jelly-roll |
| $T_{top}$ | Temperature at top centre of LIB | $K_{r3}$ | Radial thermal conductivity of positive terminal |
| $T_{bottom}$ | Temperature at bottom centre of LIB | $K_{r4}$ | Radial thermal conductivity of cap |
| $Q$ | Convective heat transfer | $K_{z1}$ | Axial thermal conductivity of negative terminal |
| $\Delta T$ | Temperature difference | $K_{z2}$ | Axial thermal conductivity of jelly-roll |
| $T$ | Temperature | $K_{z3}$ | Axial thermal conductivity of positive terminal |

# 1 Introduction

Batteries are widely used sources of energy for powering the various electronics and electrical devices. With the advancements in the field of chemistry and material science, the battery technologies evolved gradually from its version discovered by Alessandro Volta to the current lithium-ion batteries (LIB) invented by John Goodenough [1,2]. Moreover, its application in electric vehicles as an energy source has increased its eminence in the energy storage market. This is due to their high specific energy, low self-discharge rate, as well as low cost [3,4]. After all these advantages, temperature sensitivity of LIB is one of the major concerns and need to be addressed appropriately [5,6] for different ambient conditions. Even in low temperature environment, its capacity decay accelerates due to increase in internal resistance resulting into higher heat generation [7]. In addition, considerable heat is generated in the battery during the charge/discharge process as it involves endothermic/exothermic reactions [8,9], though the temperature rise is seen in both the cases. This charge/discharge process and associated parameter such as heat generation have severe impact in the specific energy, specific power, and durability of the battery [10]. On one hand, a single degree increase in the temperature above optimum operating range can reduce the performance of battery significantly [11,12]. On the other hand, even one degree of temperature heterogeneity can significantly affect the battery life [13,14]. This demands for an efficient battery cooling system to control both the temperature as well as temperature heterogeneity.

Typically, battery cooling systems (air, liquid, phase change materials) are designed to maintain the temperature with in the operating range, wherein, micro channel based liquid cooling systems are found to be more effective. According to Wang et al [15], use of nanofluids in micro channel-based liquid (deionized water) cooling systems further improved the cooling performance due to the reduction in thermal resistance by12.5–14.7%. While designing the cooling systems, temperature heterogeneity is equally important and needs to be considered. In this respect, the role of cooling approaches such as surface cooling or tab cooling becomes vital towards controlling both the temperature and temperature heterogeneity. According to a report, capacity drop of 9.2% was observed for surface cooling in comparison to 1.2 % for tab cooling of pouch cell [16]. Further, it can be pointed out that battery internal temperature is quite different from its surface temperatures. This is attributed to the internal heat generation and very low thermal conductivity of Jelly-roll [17,18]. Consequently, the time dependent temperature heterogeneity is induced that can lead to non-uniformities in current flow, hot spots, state of charge, localized particle stress and localized degradation [19]. This temperature non-uniformity is most common in cylindrical batteries [20] during its discharging or charging conditions. At high discharge rate, heat accumulates near the center of cylindrical batteries due to slow heat dissipation in radial direction compared to axial [21]. This is attributed to spiral-wound layered internal structure of the cylindrical batteries with higher thermal resistance in the radial direction than in the axial direction. Development of temperature heterogeneity is also noticed during battery charging process and its severity varies with the varying charging conditions [22]. In recent days, demand for fast-charging batteries is increasing and battery charging time is one of controlling factor for EVs growth.. These fast-charging batteries can further pose safety threat to the EVs due to the rapid temperature rise during the charging process, which is less likely for discharging process under normal driving conditions [23]. In a study, significant temperature rise (from 20 °C to 154.4 °C), was recorded at fast charging (8C rate) condition [24]. A recent study covered the aspects of temperature distribution and temperature heterogeneity for prismatic cell under fast charging (2.5 C) condition [25]. The study presented the thermal management approach for eight-cell module using parallel mini-channel based liquid cooling system. It was shown that the depth of the mini-channel has significant impact on the cooling effect (70.8%) and temperature uniformity (75.7%). At fast charging rates, the temperature uniformity can also be controlled by exchanging the coolant inlet and outlet [26]. It provided a notable perspective of cooling system design concerning to the temperature control and maintaining the temperature heterogeneity during both the charging and discharging processes. It is to be mentioned here that temperature uniformity can be measured in terms of cell maximum temperature difference (CMTD [27]. As a general trend, cooling systems design criterion is based on the philosophy of heat transfer augmentation between the batteries and surrounding, without giving due importance to the battery geometry. Battery geometry is a very important parameter that defines the charge-storage capacity, durability, and thermal behaviour of battery [28,29], hence its effect should be accounted while designing effective cooling systems. It has also been reported recently that, the failure behaviour of battery under mechanical abuse, strongly relies on the volume of active materials that changes with the geometry of battery [30]. These results substantiate the significance of geometry on thermal and structural behaviour of LIBs. Hence, geometric dependencies are required to be analysed to understand the aspects of temperature and temperature heterogeneity developed in LIBs.

According to Abada et al. [31], experimental prediction of battery internal temperature is a difficult and costly task. Also, the results were found to be less reliable in the real-time situation of a battery at different charging/discharging rate. During the operation-state, where batteries are mostly assembled as a pack and module ina vehicle or device, the scenario is much more complicated and differs drastically from the experimental prediction. Therefore, the numerical modelling garnered the need fullness to improve the understanding of thermal aspects of battery. Finite element method (FEM), lumped analysis or electrochemical methods are the popularly used modelling approach for this purpose [32-34]. However, their own limitation and assumptions restrict the prediction to match with the real value. Although, electrochemical thermal model is considered as a most effective method to model the heat generation, but the approach is very complicated and even computationally expensive. It has been observed that the electrothermal model is way a simpler and less computational as compared to the electrochemical thermal model. Thus, to develop an appropriate phenomenological model and to simulate the accurate thermal behaviour of batteries, the battery electrode dependent modelling is essential. Such models may vary with different battery technologies [35,36] with different electrode material hence computationally expensive. A comparatively less computational three-dimensional (3D) model using the equivalent circuit network (ECN) approach, with Python code is also used widely [37,38] in addition to other models [39]. However, ECN approach also lags in achieving the appropriate accuracy level. In this direction, a recent approach considering a battery-level multi-partition modelling [40] is reported to have improved accuracy while predicting battery thermal behaviour.

In view of this, the recent multi-partition modelling approach has been adopted in the present work. The research carried out in the present work uniquely is to predict the internal temperature variation with improved accuracy by using 3D multi-partition model, which realistically accounts the heat generated by different components in LIB. The present numerical analysis considers the heat generation during battery discharging process which is analogous to the vehicle driving condition that pose difficulty in in situ measurement of temperature data. Further, the present work quantitatively demonstrates the influence of three different battery geometries on the temperature heterogeneity and average temperature values. This geometry dependent heat transfer analysis would be useful while designing efficient cooling system for battery packs with different battery geometry, which is found to be unaccounted in earlier studies. The effect of cooling is examined to find out the best geometry, among the three options that can be used in electric vehicles, with minimum average surface temperature, axial temperature distribution, and radial temperature distribution for different cooling approaches. The radial convective cooling, both-tab cooling and mixed cooling are taken into the consideration with heat transfer coefficient equals to 25 W/m$^2$K (*natural convection*), 50 and 100 W/m$^2$K (*forced convection*) under different discharge rates of 1C, 2C and 3C. The natural convection condition is used to validate the FEM model. The different thermal behavioural parameters such as battery surface temperature, temperature of different battery components, axial and radial temperature difference with time for different discharge rates are numerically predicted. Subsequently, the numerical results are compared for the considered geometries. Comparative analysis of geometry-influenced cooling performances of different cooling approaches substantiated the pronounced effect of battery geometry.

## 2. Methodology

The present work considers the commercially available 18650 Li(Ni$_{1/3}$Co$_{1/3}$Mn$_{1/3}$)O$_2$LIB having nominal capacity of 2.2 Ah , nominal voltage of 3.7 V, cycle life of 1500. [40]. Three different geometriesof the considered LIBi.e., 18650 are reffered as datum geometry (DG), large geometry (LG), and small geometry (SG) which are obtained by varying the cell dimensions only.

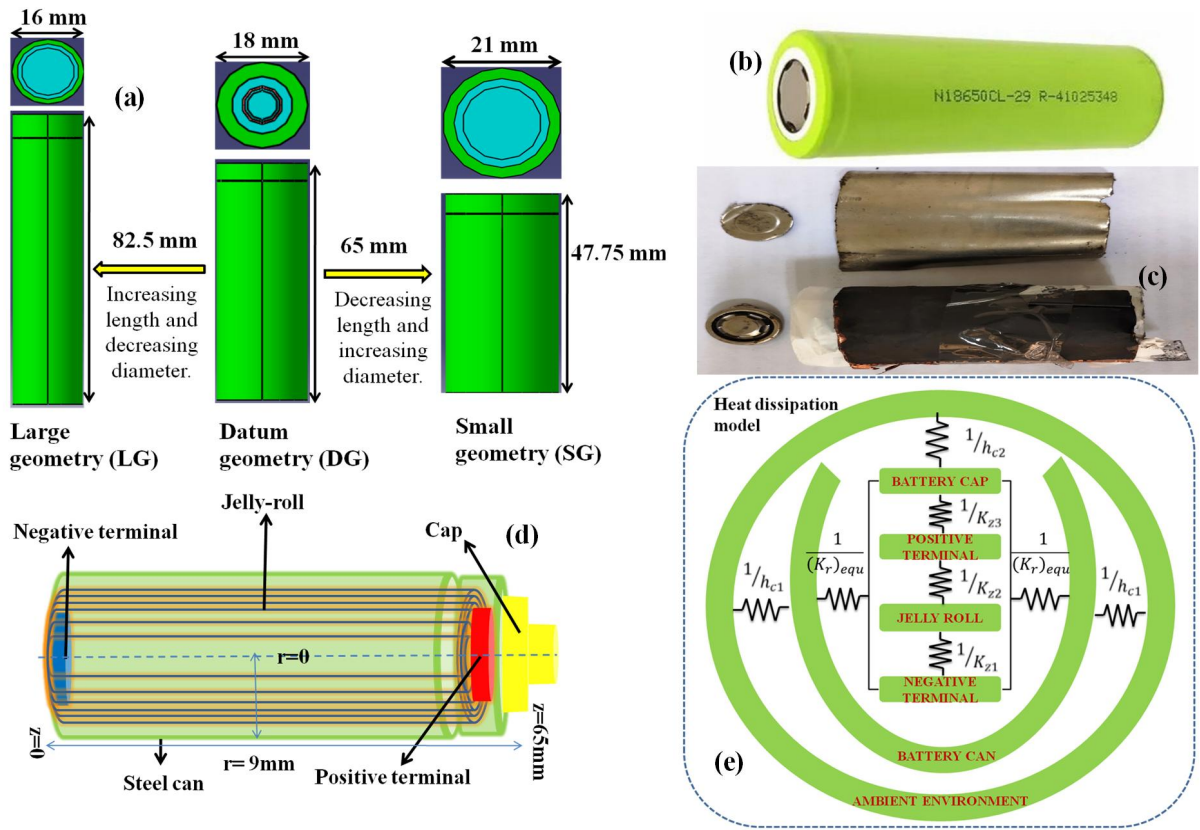

***Figure 1. Geometry and components of LIB.****(a) Top and front view of different battery geometries (LG, DG, and SG) with dimensions(b) Original picture of DG battery , (c) unassembeled picture of comercially used 18650 lithium-ion battery. (d) schematic diagram of assembeled three dimension geometry model. (e) sketch of multi-partition heat dissipation model.*

The dimensions of LG, DG and SG are illustrated in Figure 1(a). Different parts of the battery are designed separately and then assembled to form a complete battery. Figure 1(b) represents the commercially available DG and Figure 1(c) corresponds to the disassembled battery. To understand the design and assembly of different components, the battery is split into several components in accordance with the 3D multi-partition model, as presented in Figure 1(d). The various components are positive terminal, negative terminal, Jelly-roll (with cathode, anode and separator), battery cap and battery can. The present work considers the contribution of these individual components of LIB in the heat generated due to their resistance to the current flow in the resistance network during the discharging, as shown in Figure 1(e).

*2.1 Heat generation and dissipation*

The 3D multi-partition modelling consists of 3D multi-partition geometry creation, multi-partition heat generation model and multi-partition heat dissipation model. Among the components, battery can is the outermost and non-heat generating part, though participate in heat dissipation. The volume of negative terminal, positive terminal, Jelly-roll and battery cap are kept constant for all three geometries in order to maintain the same volumetric heat generation. However, the overall volume of battery for the three geometries may differ slightly after rounding off the dimensions, considering its negligible effect on the heat generation. The general and simplified form of Bernardi's expression as shown in equation (1) is considered in the present work to evaluate heat generation of different components of battery.

$$Q = I(OCP - V_{cell}) + I\left(T\frac{\partial U}{\partial T}\right) \tag{1}$$

Here, $I$ represent the charging/discharging current and $T$ is the temperature. The first term represents the irreversible component while the second term represents the reversible component. Inequation (1), the irreversible component of the heat is due to the difference between the open-circuit potential (OCP) and its operating potential ($V_{cell}$), while the reversible heat component primarily is due to the entropy change with material phase change [41]. Heat generation due to entropy change is negligible in comparison to the Ohmic heat and hence not considered here.

By replacing irreversible heat generation term $I(V - U)$ with resistance term in equation 1,

$$Q = I^2(R_p + R_o) \tag{2}$$

Here, the internal resistance of the battery (R) is segregated into ohmic resistance ($R_o$) and polarization resistance($R_p$).

The heat generation inside negative terminal ($Q_{neg}$) having internal resistance $R_{neg}$ is,

$$Q_{neg} = I^2 R_{neg} \qquad (3)$$

The heat generation equations for positive terminal and battery cap are same like equation 3. Heat generation inside the Jelly-roll ($Q_{jr}$) having internal resistance $R_{jr\_o}$ and $R_{jr\_p}$ corresponding to Ohmic and polarization resistance, respectively is expressed as below:

$$Q_{jr} = I^2 \left( R_{jr_o} + R_{jr_p} \right) \qquad (4)$$

The heat generated due to the internal resistance of the components inside the battery get transferred to the battery surface through conduction mode and then get dissipated to the external environment *via* convection and radiation. The present modelling approach considers effect of directional anisotropy of heat conduction in the LIB. The major heat interaction between the battery surface and ambient environment is mainly due to the convection since the contribution of radiation at low temperature is minimal and hence neglected. During the internal and external heat transfer, part of the heat is absorbed by the internal material itself is due to their heat affinity, which leads to the increase in the battery temperature.

Considering the same assumptions as per the battery thermal model discussed earlier [42,43],the 3D heat balance equation for DG, LG and SG cylindrical batteries can be expressed as below in equation (5):

$$\rho C_p \frac{\partial T}{\partial t} = \frac{\partial}{\partial r}\left( K_r \frac{\partial T}{\partial r} \right) + \frac{\partial y}{\partial \varphi}\left( K_\varphi \frac{\partial T}{\partial \varphi} \right) + \frac{\partial}{\partial z}\left( K_z \frac{\partial T}{\partial z} \right) + Q_{gen} \qquad (5)$$

where, $Q_{gen}$ refers to the total heat generation, which is the summation of heat generated by individual battery components, $K_r, K_\varphi$ and $K_z$ are the thermal conductivity in radial, angular and axial direction of cylindrical coordinate, respectively. $\rho$ and $Cp$ are the density and specific heat capacity of the battery components respectively, and $t$ is the time step. Here, thermal resistance model is used to express the heat interaction phenomenon between the battery and environment, and internal heat conduction between the battery components. The present work ignores the thermal contact resistance between the different internal components that contribute in heat generation. This is because of its negligible contribution (~ 3%) to heat generation for low (< 4C) charging/discharging conditions [44]. The thermal contact resistance becomes significant for high rate of charging/discharging. Hence, it may affect accuracy of prediction of heat generation and require its consideration for simulation of thermal response under high rate of charges/discharge.

The thermal network diagram shown in Figure (1e). Each battery component has different specific heat capacity and thermal conductivity. The total equivalent thermal resistance along radial and axial can be written as in equation (6).

$$\frac{1}{(R_{thermal})_{Radial}} = \frac{1}{\frac{1}{(K_z)_{equivalent}}} = \frac{1}{\frac{1}{K_{z1}} + \frac{1}{K_{z2}} + \frac{1}{K_{z3}}} \qquad (6)$$

$$(R_{thermal})_{Axial} = \frac{1}{(K_r)_{equivalent}} = \frac{1}{K_{r1}} + \frac{1}{K_{r2}} + \frac{1}{K_{r3}} + \frac{1}{K_{r4}} \qquad (7)$$

Here $(K_z)_{equivalent}$ and $(K_r)_{equivalent}$ are equivalent thermal conductivity in axial and radial direction, respectively. $K_{r1}, K_{r2}, K_{r3}, and K_{r4}$ represents radial thermal conductivity of battery cap, positive terminal, jelly-roll, negative terminal respectively, and $K_{z1} K_{z2} K_{z3}$ are axial thermal conductivity of negative terminal, jelly-roll, positive terminal respectively. Thermal resistance is the reciprocal of thermal conductivity. Similarly, $(R_{thermal})_{Radial}$ and $(R_{thermal})_{Axial}$ are the equivalent conductive thermal resistance in radial and axial direction, respectively.

It can be seen from Figure (1e) that only the battery can and battery cap are found to be participating in the heat interaction with external ambient environment through convection mode of heat transfer. Here, convective heat transfer can be represented as thermal convective resistance. Therefore, external equivalent convective thermal resistance are $\frac{1}{h_{c1}}$ and $\frac{1}{h_{c2}}$, where $h_{c1}$ and $h_{c2}$ are the convective heat transfer coefficient for battery can and battery cap ,respectively. Since, the equivalent thermal conductive and convective resistance relies on the geometry of cylinder (length and diameter), it is expected to observe different rates of heat conduction and convection resulting into different cooling performance DG, SG and LG. In accordance with the thermal multi-partition model, the total electric resistance of the LIB can be expressed as below:

$$R_{total} = R_{pos} + R_{neg} + R_{cap} + R_{jr\_o} + R_{jr\_p} \qquad (8)$$

where, $R_{total}$ represent the total electric resistance of the battery, $R_{jr\_o}$ and $R_{jr\_p}$ denotes the Jelly-roll resistance. $R_{pos}, R_{neg}$ and $R_{cap}$ are the resistance of positive terminal, negative terminal, and battery cap, respectively. As in the present study, the effect of geometry change is being analysed without changing the total volume of battery and battery components, the total volumetric heat generation $(Q_{vol})_{Total}$ for the different types of geometry is considered to be same as expressed in equation (9).

$$((Q_{vol})_{Total})_{DG} = ((Q_{vol})_{Total})_{LG} = ((Q_{vol})_{Total})_{SG} \qquad (9)$$

Figure 2(a) represents the reproduced data of the experiments [40] illustrating the variation of resistance of individual components with state of charge (SOC). It is observed that the resistance of the cap, positive terminal, and negative terminal are constant, while Jelly-roll resistance varies significantly with SOC. Using equation (4), the variation of volumetric heat generation for Jelly-roll at various discharge rate is calculated with respect to the depth of discharge (DOD), as shown in Figure 2(b). The nominal currents for DG are 2.2, 4.4 and 6.6 A at the discharge rate of 1C, 2C, and 3C, respectively. A rapid rise in the heat generation of Jelly-roll resistance is observed after around 80 % of DOD. During the discharge process, exothermic reaction takes place, due to which the rise in temperature and temperature heterogeneity is higher compared to that observed in charging condition [6, 19]. And close to the end of discharge (after around 80 % of DOD in most of the cases), the battery capacity drops significantly leading to a decrease in battery voltage. In order to maintain the constant power, current increases, consequently, heat generation rate increases.

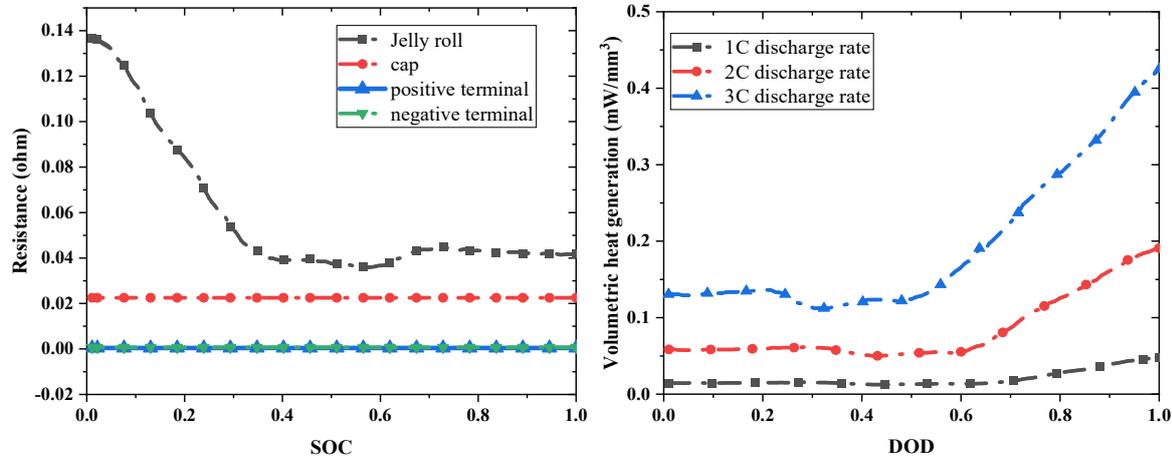

***Figure 2. (a) Variation of internal resistance of battery components*** *with state of charge (SOC) [40](b) Variation of volumetric heat generation of Jelly-roll with the depth of discharge (DOD) at 1C, 2C and 3C discharge rate.*

The volumetric heat generation is more impactful in governing the battery temperature distribution as battery consists of different components and every component have different volume. The role of individual battery components on the volumetric heat generation is well understood by comparing the heat generation, volumetric heat generation, resistance, and volume for each part, as listed in table 1. Although the volumetric heat generation of negative terminal is highest, the battery cap attains the maximum temperature during charging/discharging process. Consequently, the battery cap is the major contributor towards developing temperature heterogeneity within the battery. On the contrary, the temperature rise for the positive terminal and negative terminal are less compared to that of Jelly-roll and battery cap despite having higher amount of volumetric heat generation. This is due to the lesser volume of those components in comparison to Jelly-roll and battery cap.

**Table 1.** *Volume and resistance of each part of battery, heat generation and volumetric heat generation of each part of battery at 1C, 2C, and 3C discharge rate [40].*

| Components | Discharge rate | Cap | Positive terminal | Negative terminal | Jelly roll |
|---|---|---|---|---|---|
| Volume (mm³) | - | 739.61 | 10 | 4.780 | 14100 |
| Resistance (ohm) | - | 0.0225 | 0.000442 | 0.000663 | Variable |
| Total heat generation (mW) | 1C | 108.9 | 2.139 | 3.209 | 663.18 (at DOD 1) |
|  | 2C | 435.6 | 8.557 | 12.836 | 2671.34 (at DOD 1) |
|  | 3C | 980.1 | 19.253 | 28.880 | 5951.40 (at DOD 1) |
| Volumetric heat generation (mW/mm³) | 1C | 0.147 | 0.214 | 0.671 | 0.04737 (at DOD 1) |
|  | 2C | 0.588 | 0.856 | 2.685 | 0.19081 (at DOD 1) |
|  | 3C | 1.325 | 1.925 | 6.042 | 0.4251 (at DOD 1) |

Unlike positive terminal, negative terminal, and cap, Jelly-roll's heat generation varies with SOC/DOD significantly at lower SOC or higher DOD. This happens because, close to the end of discharge, the battery capacity drops significantly leading to decrease in battery voltage. In order to maintain constant power, current increases, consequently, heat generation rate increases for Jelly-roll.

## 2.2 *Thermo-physical properties*

The thermo-physical properties of each battery component, as needed to perform multi-partition thermal modelling, are taken from the literature [40] and listed in Table 2.

**Table 2.** *Thermo-physical properties of different components used in battery [30, 40, 41]*

| Components | Material | Density (kg/m³) | Heat capacity | Thermal conductivity |
|---|---|---|---|---|

|  |  |  | (J/kg K) | (W/m K) |
|---|---|---|---|---|
| Positive terminal | Aluminum | 2719 | 871 | 202.4 |
| Negative terminal | Nickel | 8900 | 460.6 | 91.74 |
| Jelly-roll | Electrode &Separator | 2440 | 1210 | $K_r$=1.1, $K_z$=$K_q$=12.5 |
| Cap | Polycrystalline titanate ceramics (PTC) | 3455 | 565.5 | 30 |
| Battery can (Steel) | Steel | 8030 | 502.48 | 16.27 |

Conduction and convection are the major thermal phenomenon responsible for the heat transfer and cooling in battery. Convective heat transfer is a surface phenomenon, it majorly depends on the surface area exposed to the environment, ambient environmental condition, velocity of fluid, and temperature difference between body and environment [45,46]. By testing all these values experimentally, Nusselt number can be found out and heat transfer coefficient ($h$) can be obtained using the following equation (10):

$$h = \frac{q}{\Delta T} \tag{10}$$

Where, $q$ is convective heat transfer per unit area and $\Delta T$ represent temperature difference between body surface and environmental fluid [47].

## 3. Finite element modelling

Finite element modelling of the LIB geometry is done with quad elements. Total number of 26876 elements are considered after performing the grid independency test to achieve the converged results of variation in average battery temperature. Transient thermal analysis is performed in ANSYS considering the equation (5), with the boundary conditions for conduction and convection heat transfer as below:

$$At \ r = 0, \frac{\partial T}{\partial r} = 0; \ At \ r = r_o, \ -K_r \frac{\partial T}{\partial r} = \ h_r \ (T - T_o) \tag{11}$$

$$At \ z = 0, \frac{\partial T}{\partial z} = 0; \ At \ z = z_o, -K_z \frac{\partial T}{\partial z} = h_z(T - T_o) \tag{12}$$

Here, $T$ is the temperature at any instant, $T_o$ is the ambient temperature, $r_o$ and $z_o$ are the maximum radius and length when measured from the centre and base of the considered LIB, respectively. Centre and base of LIB corresponds to $r = 0$ and $z = 0$ respectevily, as shown in figure 1(d). Initial condition as per the equation (13) is employed to perform the analysis:

$$T(r, \varphi, z, 0) = T_o \tag{13}$$

The numerical results using the considered multi-partition modelling approach is validated with the existing experimental work for the average surface temperature [40]. The referred experiment was conducted in a temperature-controlled environment. The batteries were discharged at different current rates (1C, 2C and 3C) to analyse the thermal behaviour. The BTS-20 battery test system was used to monitor the current, terminal voltage and the cell surface temperature. The battery surface temperature was measured at different discharge rate (1C, 2C, and 3C) using thermocouples. In addition, a commercial thermal infrared imager was used to characterize the battery surface temperature distribution at three discharge current rates (1C, 2C and 3C). The temperature data were recorded at every 10% of DOD, where 0% denotes fully charged, and 100% is fully discharged. The validation of the present approach is done using DG corresponding to radial cooling at 1C, 2C, and 3C discharge rate with heat transfer coefficient equals to 25 W/m²K and 13 °C ambient temperature. Figure 3a shows the comparison of the experimental and numerical results. The error between the experimental and numerical predictions are calculated for 1C, 2C and 3C, wherein the average error is close to 2.2 % and the maximum error is around 5.6 % for 2C case. Figure 3(b)is the error corresponding to 2C rate to display the variation of percentage error with time.

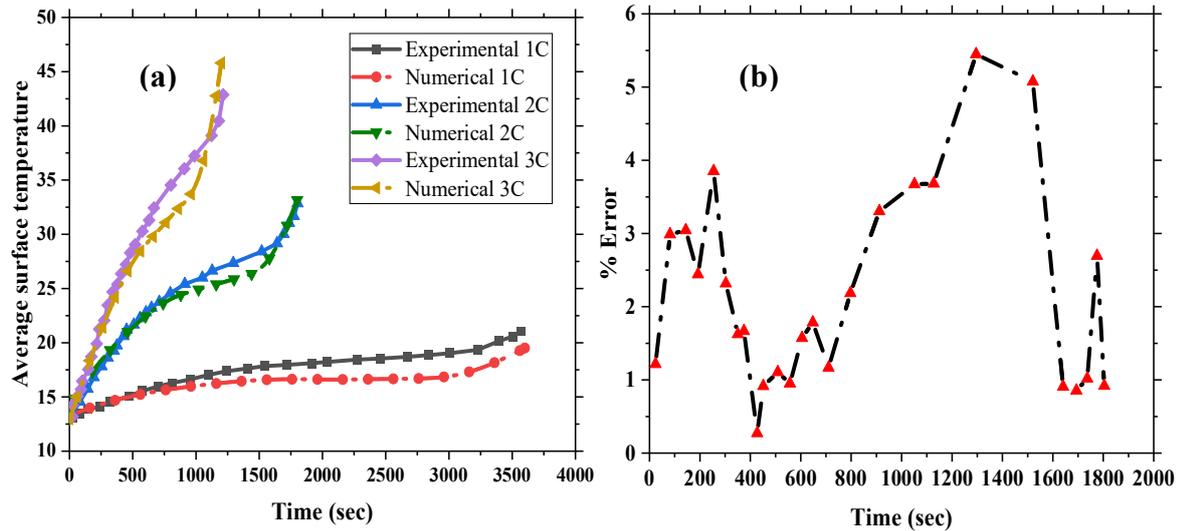

***Figure 3. (a) average surface temperature with time (experimental [40] and numerical) for lithium-ion 18650 battery, (DG) at 1C, 2C, and 3C discharge rate, (b) Variation of percentage error between experimental and numerical prediction for 2C discharge rate.***

## 4 <u>Results and discussion</u>

Influence of geometry on cooling performance, in terms of temperature heterogeneity and average temperature for radial, both-tab cooling and mixed cooling is demonstrated herein. The schematic of the considered cooling approaches are illustrated in figure 4. Considering the radial cooling approach, the curved surface of cylindrical LIB is responsible for the heat interaction from surrounding. It is the most used technique. It significantly reduces the average battery temperature. On the other hand, the tab cooling uses the top and bottom surfaces (i.e., positive, and negative terminals) for heat transfer. This technique is less efficient in controlling the average battery temperature but provide best radial thermal uniformity. However, the practicality of both-tab cooling is challenging because of the presence of battery connections and presence of BMS connection wires. Mixed cooling combines the beneficial features of both the approaches where all the surfaces i.e., curved, top and bottom surface, actively participates in heat transfer process to surrounding. It provides both the lower average battery temperature as well as better thermal uniformity in radial and axial directions. The detailed analysis of these cooling approach is discussed in the sections 4.1, 4.2 and 4.3 covering the aspects of the effectiveness of these approaches in case of forced cooling conditions.

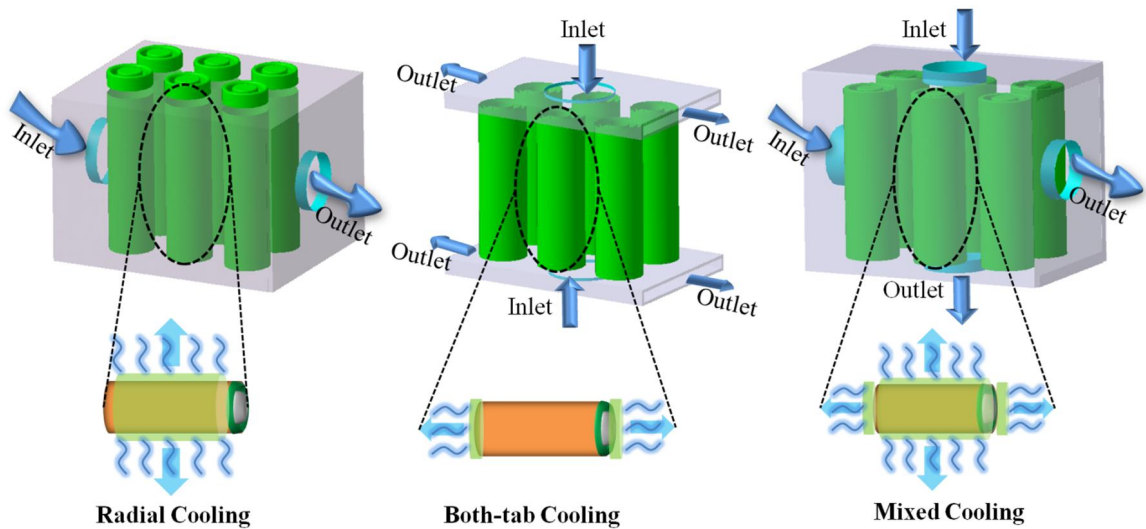

*Figure 4. Schematic representation of radial, both-tab and mixed cooling approaches.*

The quantitative comparison of the time dependent average surface temperature and temperature distributions is presented in section 4.4 for the considered forced cooling conditions corresponding to the geometries considered.

### 4.1 *Effect of radial cooling*

The heat transfer coefficient of 50 W/m$^2$K is considered to analyse the effect of radial cooling. Curved surface is considered as a convective heat transfer area (excluding top and bottom cross-sectional area). To probe the insights on the geometry-dependent thermal behaviour of battery, the results for axial and radial temperature difference for radial cooling is discussed herein. It is noteworthy to say that that the experimental prediction of spatio-temporal distribution of battery's internal temperature is complicated and require the use of micro thermocouple inside the battery which may affect the battery performance [48]. Whereas, it becomes easy to accurately estimate the same using 3D multi-partition model.

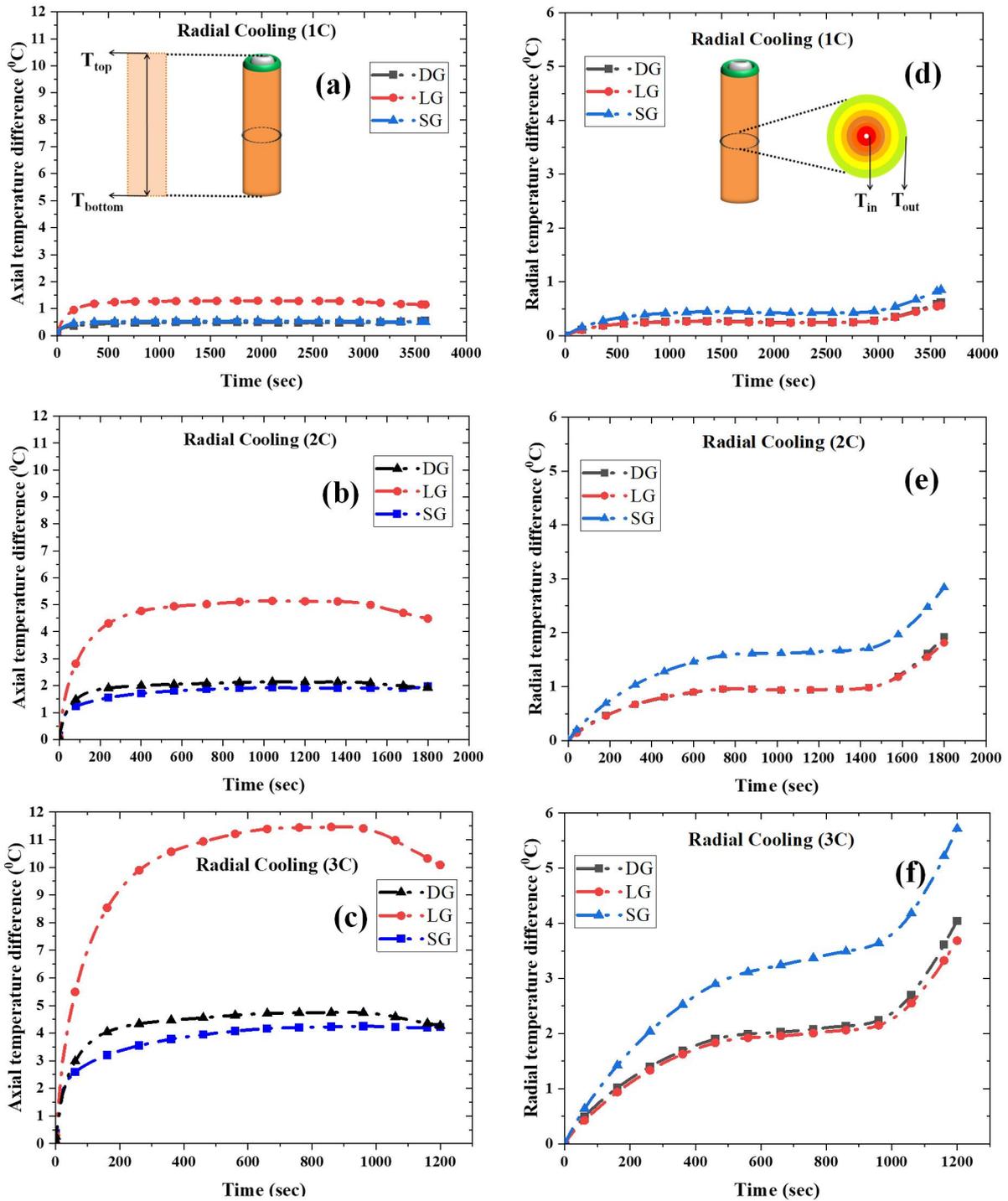

**Figure 5. Geometry dependent (LG, DG and SG) temperature** *difference* **variations with time for radial cooling at 50 W/m²K:** *Axial temperature difference for (a) 1C, (b) for 2C, and(c) 3C discharge rate; Radial temperature difference for (d) 1C, (e) 2C, and (f) 3C discharge rate.*

As shown in the inset figure 5, $T_{in}$, $T_{out}$, $T_{top}$ and $T_{botttom}$ are temperature probes at the inner most surface of Jelly-roll, at outer surface, on the battery cap and at bottom of battery respectively. These temperature values are used to estimate temperature difference in both the radial and axial directions with respect to time. Here, axial temperature difference is the difference between $T_{top}$ and $T_{bottom}$ and radial temperature difference is the difference between $T_{in}$ and $T_{out.}$ The axial temperature difference versus discharging time for 1C, 2C, and 3C discharge rate is shown in Figure 5 (a-c). Amongst the LG, DG, SG, the axial temperature difference is found to be maximum for LG as it has the maximum axial length which leads to maximum axial thermal resistance, and SG have minimum axial temperature difference due to its minimum axial length, irrespective of discharge rate. On the other hand, Figure 5 (d-f) substantiates that the radial temperature difference is minimum *for LG* and maximum for SG under radial cooling, irrespective of discharge rate. This can be attributed to the small radial length for LG and longer radial length of SG.

It has been shown previously that non uniform temperature distribution can lead to inhomogeneous current flow in a parallel string of cells in a battery pack [49]. According to Hunt and coworkers [16], different layers of LIB are analogous to several number of small cells of different temperature connected in parallel, hence these small cells become hot spots with different temperature. During discharge the cell at higher temperature have less resistance to the charge flow, hence becomes source of higher current, which subsequently increases the temperature further. This inhomogeneous current flow leads to further increase in heat generation which is proportional to the square of the current. As a result, with charge-discharge cycles, localised electrode and separator degradation may occur leading to early battery failure due to the possibilities of internal short circuiting. Considering this, it is desired to keep the value of temperature difference to minimum for longer cycle life and to mitigate the risk of early battery failure due to internal short circuiting. In other way, smaller is the temperature difference, more effective is the thermal management considering cycle life of a battery.

Among the DG, LG and SG, at 3C discharge rate, figure 5 shows that SG exhibits *minimum axial temperature* differenceof4.22 °C, while LG shows *minimum radial temperature* difference of 3.68 °C, after 1200 sec. It is imperative to mention that radial temperature heterogeneity is more degrading as it involves multiple layers of Jelly-roll as heat generating source that behaves as smaller cells with higher current. This is in contrasts with the less severe degrading effect of axial temperature heterogeneity, which involve single layers as source of heat as smaller cells arranged in axial direction.

*4.2 Effect of both-tab cooling*

In both-tab cooling, top and bottom cross-sectional area is considered as convective heat transfer area with zero heat transfer in radial direction. Therefore, the area available for heat transfer is less in both tab-cooling compared to that for radial counterpart. Consequently, in both-tab cooling, the difference in the rate of heat generation and its release to ambient environment causes heat accumulation and rapid temperature rise within the battery. From the Figure 6 (a-c), it is inferred that the *axial temperature* difference *is minimum* for SG and maximum for LG. On the other hand, Figure 6 (d-f), shows that the *radial temperature difference is minimum for LG* and maximum for SG. With both-tab cooling at 3C discharge rate, the *minimum axial temperature difference* is 7.9 °C for SG and *minimum radial temperature difference is 0.5 °C* for LG after 1200 sec.

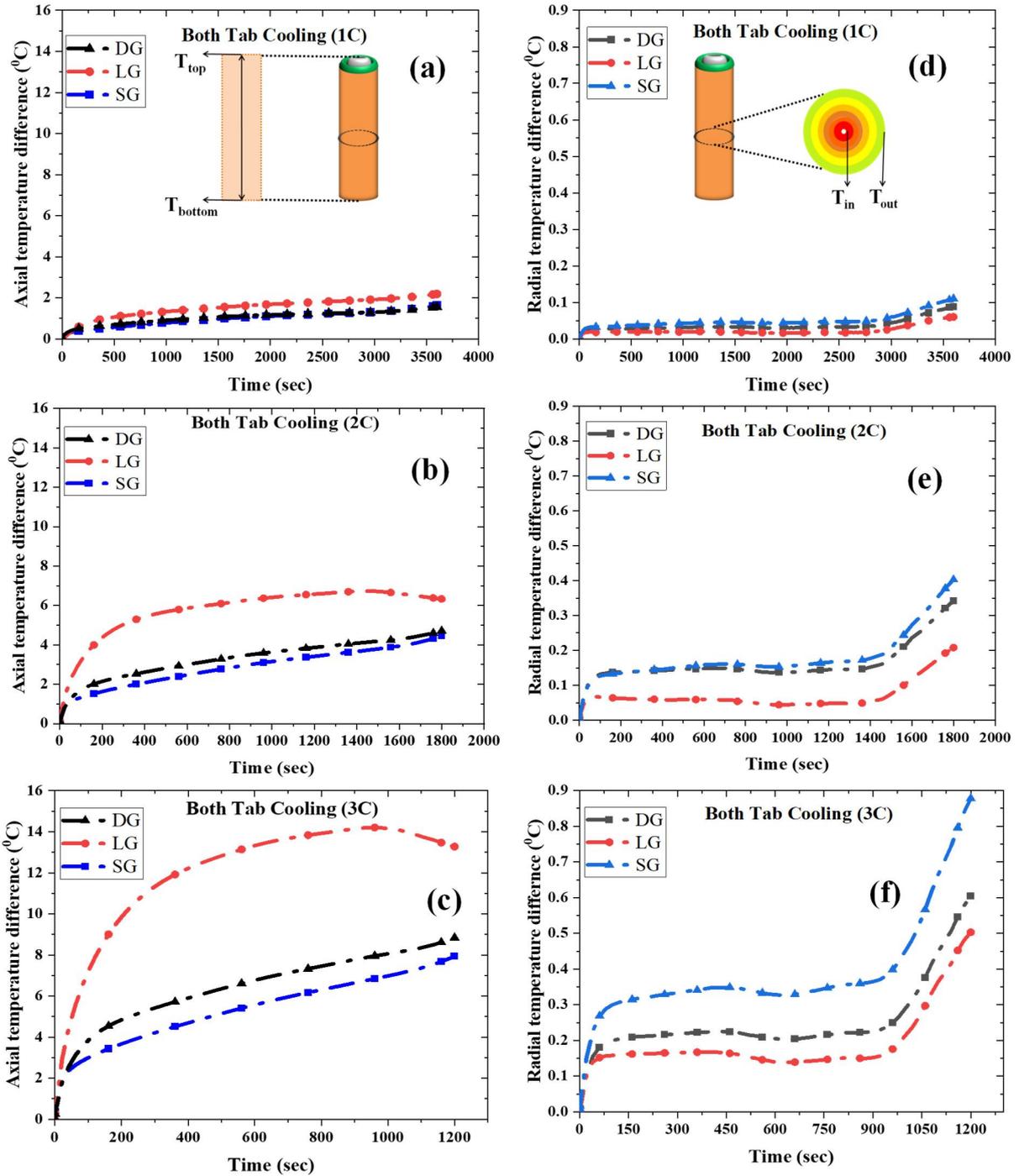

*Figure 6. Geometry dependent (LG, DG and SG) temperature difference variations with time for both-tab cooling at 50 W/m²K:* Axial temperature difference for (a) 1C, (b) 2C, and (c)3C discharge rate; Radial temperature difference for (d) 1C, (e) 2C, and (f) 3C discharge rate.

Considering the radial temperature difference having severe degrading effect, both-tab cooling approach is found to be more effective as it leads to significantly reduced the temperature heterogeneity of 0.60, 0.50 and 0.87°Cat 3C rate. It is worth mentioning that, the axial temperature difference in both-tab cooling (7.9 °C for) is more than the radial cooling (4.22 °C) with positive terminal being at higher temperature. This is because of higher Ohmic heat generation of the battery cap, though, thermal resistance in the radial direction is greater due several layers of Jelly-roll wrapped together with electrolyte salt and separator. Despite of higher battery temperature, through both-tab cooling approach, temperature uniformity in radial direction is better for all types of geometries compared to other cooling approaches.

*4.3Effect of mixed cooling*

In radial cooling of 18650 cylindrical LIB (DG), the curved area actively participated in heat interaction with the surrounding. The heat transfer takes place mostly in direction perpendicular to the layers of jelly-roll. This results in higher radial temperature heterogeneity compared to both-tab cooling, which promotes heat flow is along the axis of the battery. Mixed cooling combines the features of radial and both-tab cooling wherein battery's total surface area (curved and tab area) participates in heat transfer compared to both-tab and radial. This cooling technique helps in reducing the battery temperature to the minimum. In the mixed cooling approach, the geometry dependent axial and radial temperature difference follows the same trends as seen in other two cooling approaches. Irrespective of the discharge rates, figure 7 (a-c) reveals that LG exhibited a higher axial temperature difference with maximum value of 9.70 °C while it is *minimum* for SG (5.80 °C) at 3C, after 1200 sec. From the figure 7 (d-f), the *radial temperature difference is minimum* (3.40 °C) for LG and maximum for SG (5.17°C) at 3C rate.

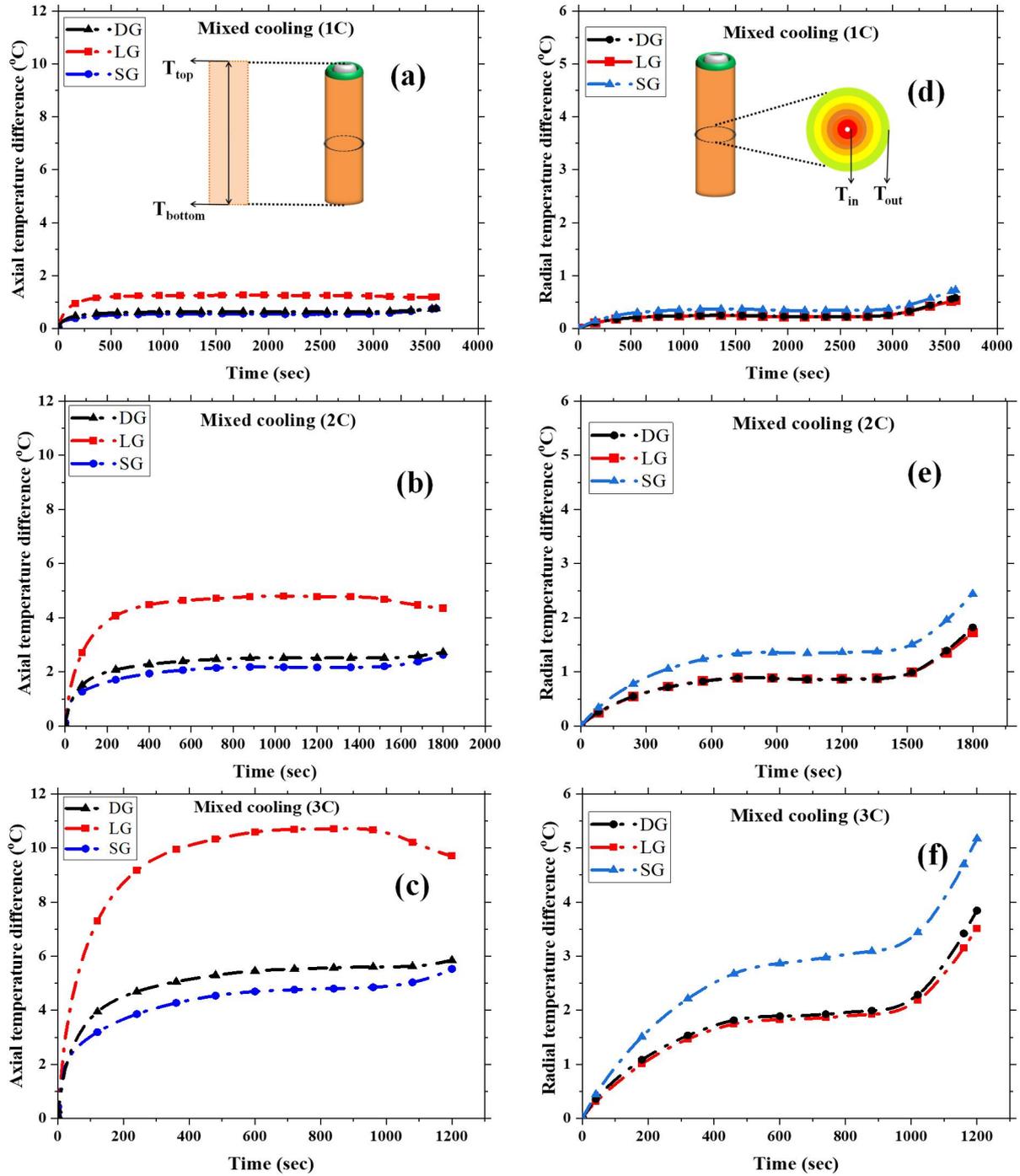

*Figure 7. Geometry dependent (LG, DG and SG) temperature difference variations with time for mixed cooling at 50 W/m²K: Axial temperature difference for (a) 1C, (b) 2C, and (c) 3C discharge rate; Radial temperature difference (d) 1C, (e) 2C and (f) 3C discharge rate.*

It can be inferred now that the geometry dependent temperature variation under mixed cooling approach combines the effect of radial and both-tab cooling approach. This result also indicates that the mixed cooling approach could be an adoptable cooling approach amongst the radial and both-tab cooling.

Thermal behaviour of LIBs discussed in section 4.1, 4.2 and 4.3 revealed that the evolution of temperature distribution with time occurs in three stages. Radial temperature distribution, shown in figure 5-7, can be categorized into primary, secondary and tertiary stages based on the percentage depth of discharge (DOD). In the primary stage (0% to 25% DOD), the rate of temperature rise is fast, which is due to slower heat dissipation due to negligible difference in battery surface temperature and ambient temperature. Thereby, the generated heat remains inside with its rapid accumulation. In secondary stage (25% to 75% DOD), the rate of temperature rise is comparatively slower, which is due to the relatively high difference in battery surface temperature and ambient temperature. In addition, reversible entropic heat generation decreases during this stage [18]. Combined effect of the two factors brings down the rise of radial temperature difference. In tertiary stage (75% to 100% DOD), a notable rise in the temperature at very high rate is depicted. Such rapid rise in the third stage is ascribed to the rapid rise in Jelly-roll resistance near the end of discharge. It is at this stage, both irreversible and reversible heat generation increases dramatically [17,18]. Referring to Figure 5-7, the curves for the axial temperature difference can also be divided into three stages i.e., primary, secondary, and tertiary. The tertiary being like the secondary stage, or its drooping behaviour characterises the higher axial conductivity compared to radial as stated previously. This, in turns, leads to the variation in the quantity $T_{top}$ - $T_{bottom}$, such that, it becomes negative. It points out that, the temperature of the negative terminal ($T_{botttom}$) rises in the third stage compared to that of positive terminal or cap ($T_{top}$) due to the lesser available area of negative terminal for convective heat transfer to ambient environment.

### 4.4 *Effect of heat dissipation on the average battery temperature*

The present section focuses on the influence of changes in the heat transfer rate on the outcome of different cooling approaches corresponding to the LG, DG and SG-type batteries. The influence is characterized by the drop in the average battery temperature.

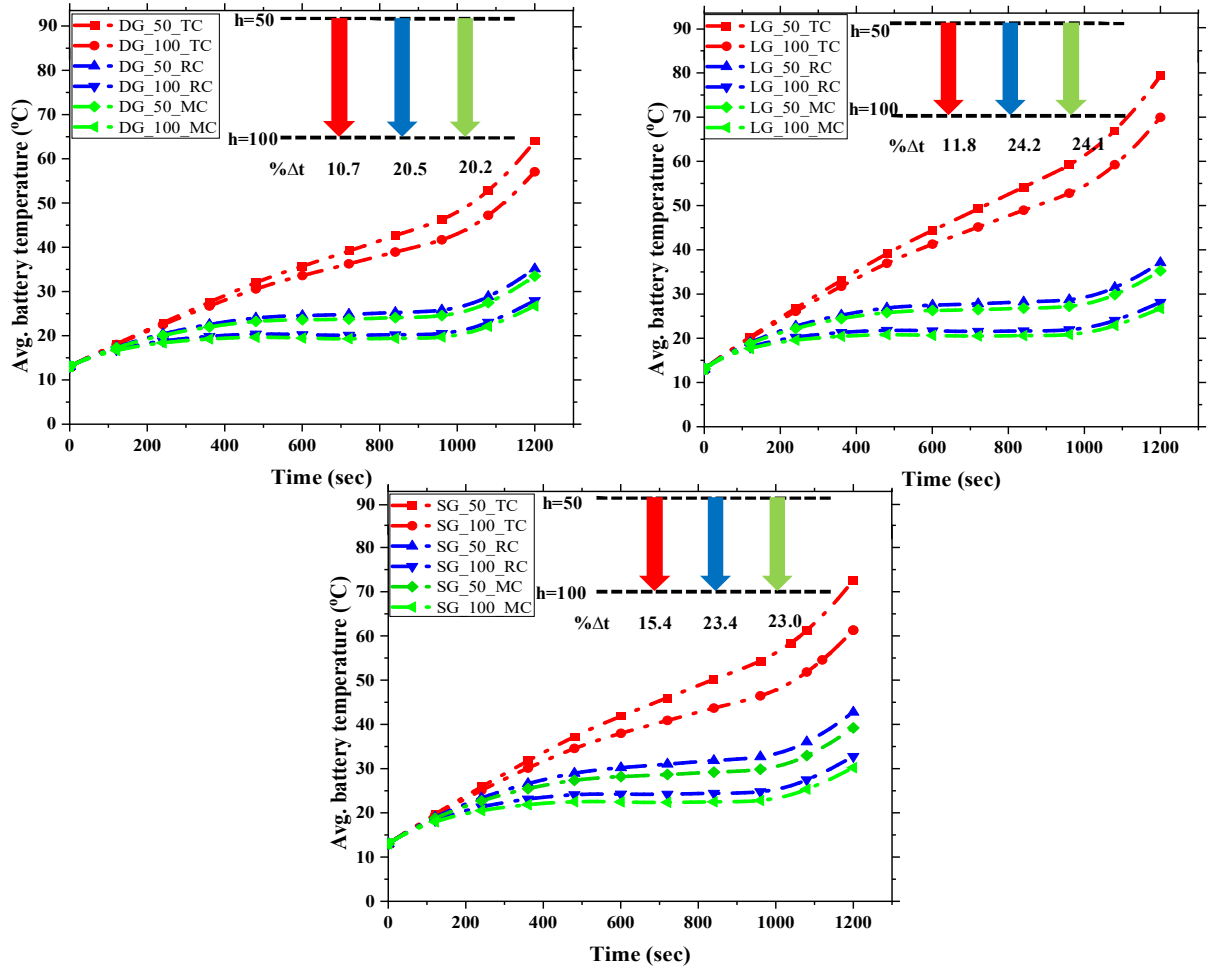

***Figure 8. Evolution of average battery temperature*** *with time for both-tab cooling (TC), radial cooling (RC) and mixed cooling (MC) at 3C discharge rate corresponding to h= 50 and 100 W/m²K for (a) DG, (b) LG and (c)SG*

Figure 8 (a-c) represents the average battery temperature with time at 3C discharge rate for DG, LG, and SG batteries for different cooling approach, with50 and 100 W/m²K. The average battery temperature curve displays similar trend with three phases as that of radial temperature difference characterizing the rate of evolution. The results in figure 8 manifests that the average battery temperature for both-tab cooling is significantly high compared to radial and mixed cooling for any battery geometry. In the case of LG, the average battery temperature approached 83.46°C for h=50 W/m$^2$K, which is far above the optimum operating temperature of the battery (60°C). This necessitates the use of forced convective cooling with a higher value of heat transfer coefficient for tab cooling of LG type batteries. Alternatively, average battery temperature for DG and SG batteries are well within the operating temperature range for radial and mixed cooling.

Radial cooling being widely adopted cooling approach, displays maximum percentage drop of 24.2% in the average battery temperature for LG in contrast to DG (20.5 %) and SG (23.4 %) with the increase in '$h$' from 50 to 100 W/m$^2$K. On account of this finding, radial cooling can be suggested for LG as per the average battery temperature criterion.

Considering the both-tab cooling approach, the maximum percentage drop in the average battery temperature with the change in '$h$' is 15.5% for SG in contrast to DG (10.7 %) and LG (11.8%). The reason for such maximum drop in SG batteries can be attributed to the fact that, its available area for convective heat transfer is maximum among the other geometries for both-tab cooling. The higher drop in average battery temperature indicates the suitability of the considered forced cooling approach for proposed battery geometry, i.e., radial cooling (forced) is good for LG batteries and both-tab cooling is good for SG.

In case of mixed cooling, the maximum percentage drop in the average battery temperature with the change in '$h$' is 24.1% for LG in contrast to DG (20.2 %) and SG (23.0%). In summary, the mixed cooling approach is found to be beneficial in terms of controlling the average battery temperature as it combines the features of radial and tab cooling. From figure 8, it is inferred that there is a profound effect of forced cooling at higher convective heat transfer rate. The effect is substantiated by the reduction in average battery temperature ranging from 13.5% to 25% for the considered cooling approaches. Contrarily, simultaneous rise in the temperature difference inside the battery is observed. In order to further demonstrate the geometry dependent temperature difference for different cooling approaches, figure 9 shows the battery internal temperature distribution at the end of discharge.

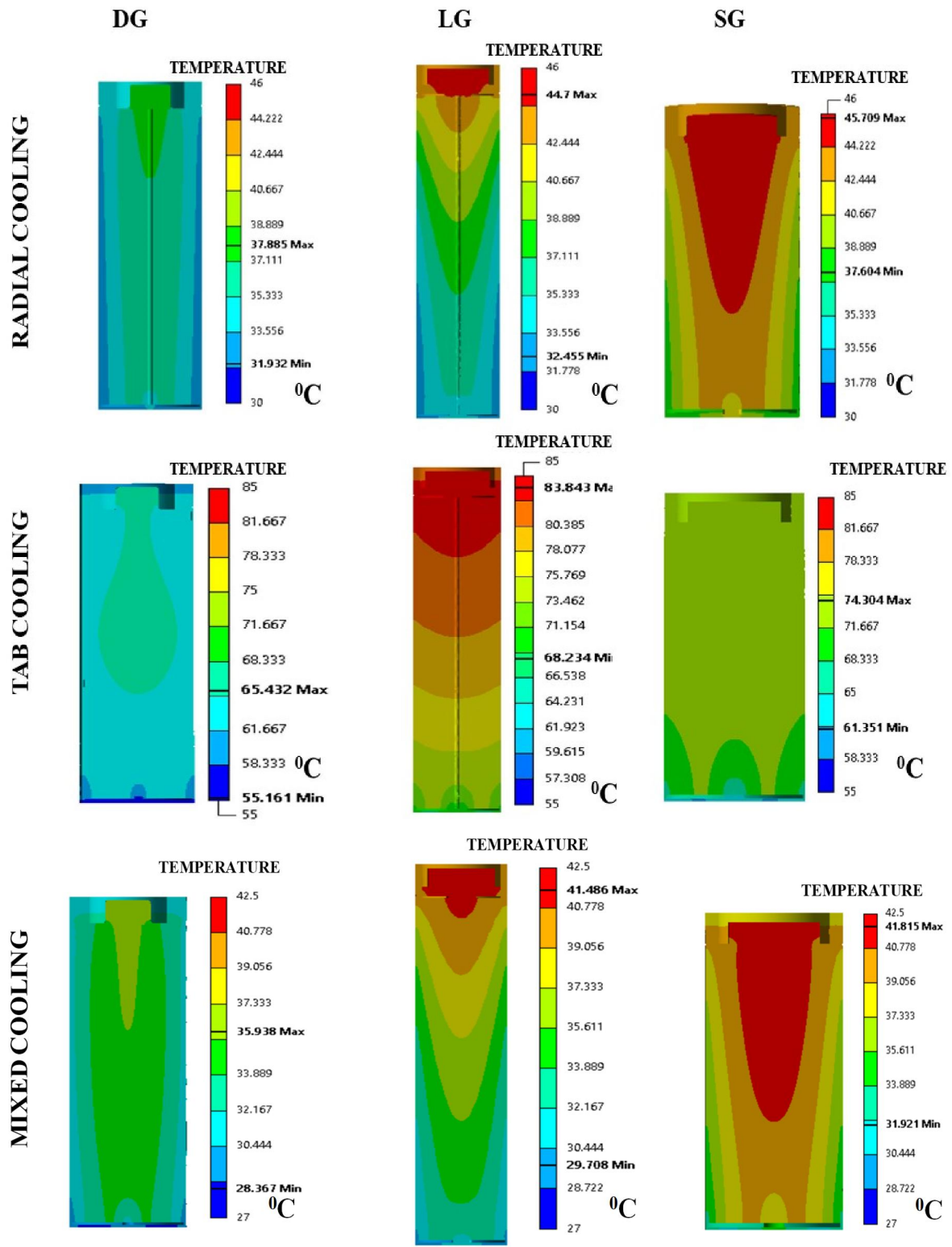

*Figure 9.* *Temperature distribution for different battery geometries (DG, LG and SG) under radial, both-tab and mixed cooling approachat3C discharge rate.*

From the Figure 9, the maximum average battery temperature of 45.70 °C is obtained for SG under radial cooling and 83.46 °C for LG under both-tab cooling at 3C discharge rate. It is also noticed that average battery temperature is minimum for DG with 35.65 °C under mixed cooling, while it is nearly equal for SG (41.4 °C) and LG (41.8 °C).

The effect of different cooling approach on different battery geometries is summarized in Figure 10 by presenting the comparative thermal behaviour for '$h$' 50W/m²K, at 3C discharge rate. The curved surface area of SG is minimum, which leads to reduced convective heat transfer, and the radial distance is maximum, which leads to increased conductive thermal radial resistance amongst all types of geometry. Due to this reason, the accumulation of heat for SG battery is high, resulting into its maximum surface temperature for radial cooling approach. Figure 10 (a-c) compares the cooling performance based on radial temperature difference, wherein LG outperforms compared to DG and SG for any of the three cooling approaches considered. Here LG with both-tab cooling shows the minimum radial temperature difference of 0.5 °C. In both tab cooling, radial temperature difference ($T_{in} - T_{out}$) is quite less compared to that of radial cooling. Also, in both-tab cooling, the convective heat transfer area is minimum which results in slow heat transfer to ambient environment. Due to higher heat conduction in axial direction, heat flux is more in axial direction compared to radial. Consequently, heat accumulates at faster rate than it dissipates causing the higher average battery temperature for both-tab cooling approach. Figure 10 (d-f) compares the cooling performance based on axial temperature difference, wherein SG is the best option among the three geometries with 4.22 °C after 1200 s for radial cooling.

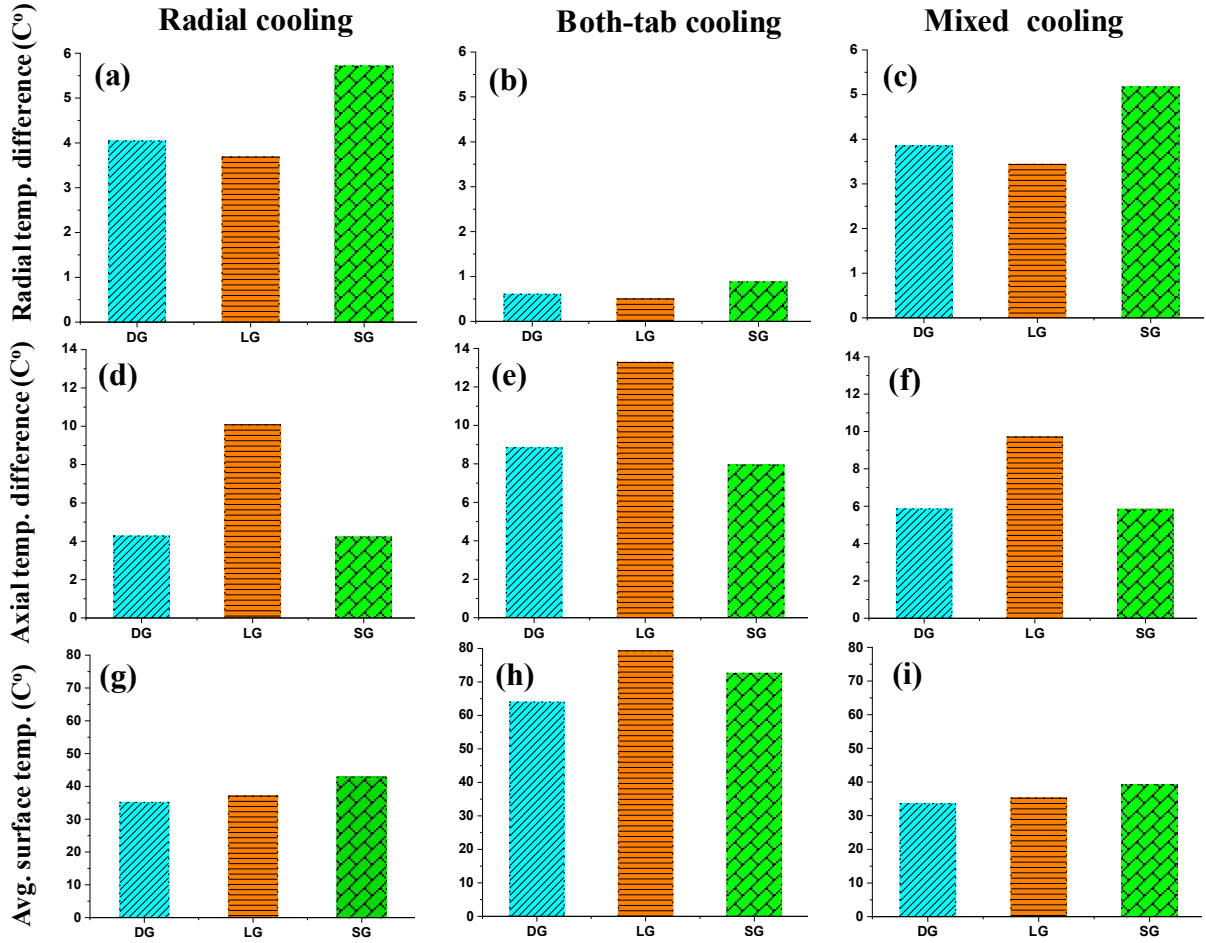

***Figure 10.*** *Temperature values of different thermal behaviour parameter at the end of discharging time for 3C discharge rate for DG, LG, and SG batteries under different cooling approaches for heat transfer coefficient of 50W/m²K.*

Figure 10 (g-i) shows the comparative performance of different cooling approaches based on the average battery temperature that governs the operating efficiency of LIBs. Accordingly, with mixed cooling approach, the best cooling performance is observed for DG compared to the LG and SG batteries. The final average temperature of DG is found to be 33.5 °C after 1200 seconds. In view of the thermal uniformity inside the LIBs, which is always being a concern for its life and reliability, its value can be quantified in terms of temperature standard deviation. As mentioned in figure 10, the radial temperature difference is minimum in LG battery irrespective of cooling techniques, therefore, it becomes important to understand and compare thermal uniformity inside the LIBs. Hence, temperature standard deviation is introduced to evaluate the thermal uniformity achieved for the different cooling approaches considered.

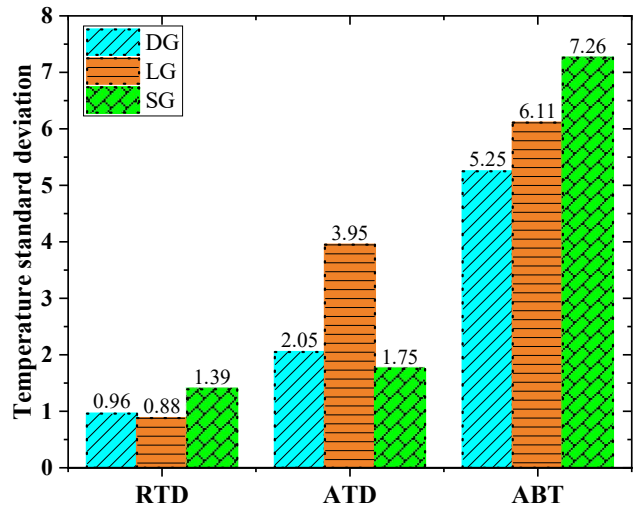

***Figure 11***. *Temperature standard deviation for different battery geometry under mixed cooling technique corresponding to h= 50W/m²K at 3C discharge rate.*

Figure 11 presents the thermal uniformity of DG, LG, and SG under different considered thermal parameters for mixed cooling approach. From the figure, it is inferred that the radial temperature difference (RTD) is most uniform for LG, axial temperature difference (ATD) is most uniform for SG, and ABT is minimum and most uniform for DG battery. Similar outcome for RTD, ATD and ABT is observed for the other considered cooling approaches, though the temperature standard deviation values are different.

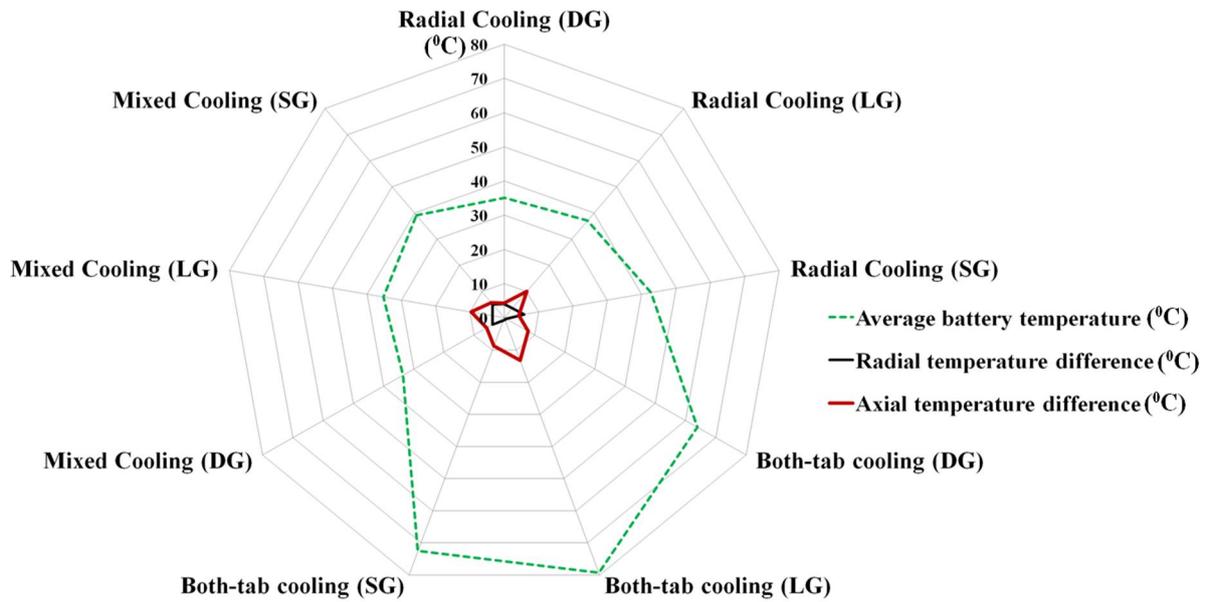

***Figure 12****. Geometry dependent cooling performance of different cooling approach corresponding to axial, radial temperature difference or average battery temperature (℃) based design criteria.*

Thus far, it is evident that the geometry is an important parameter, which governs the operating performance as well as cycle life of LIBs. Hence, it demands for choosing appropriate battery design criterion based on cooling approach with due importance to the geometry of the battery. In this connection, Figure 12 presents a simplistic strategy that may be adopted while designing a battery according to the cooling approach or vice-versa. It illustrates the approach that can be adopted for efficient cooling of different battery geometries, considered in the present work.

# 5 Conclusions

A three-dimensional multi-partition thermal model with finite element method is adopted in the present study to investigate the geometry dependent thermal behaviour of lithium-ion battery under different types of cooling approaches. Different types of geometries are obtained by changing the longitudinal and diametric length of reference/datum geometry i.e., lithium-ion 18650 cylindrical battery. The strong influence of battery geometries on the cooling performance is demonstrated. The important conclusions are listed as below:

a.  One hand, for thermal management systems of LG batteries, both-tab cooling is recommended if the design criterion is based on minimum radial temperature difference. On the other hand, DG is appropriate with mixed cooling approach as it shows minimum battery temperature. SG is appropriate with radial cooling approach and gives minimum axial temperature difference. It is to be highlighted that

    $(ATD)_{Radial\ cooling} < (ATD)_{Mixed\ cooling} < (ATD)_{Both\text{-}tab\ cooling}$,

    $(RTD)_{Both\text{-}tab\ cooling} < (RTD)_{Mixed\ cooling} < (RTD)_{Radial\ cooling}$ and

    $(ABT)_{Mixed\ cooling} < (ABT)_{Radial\ cooling} < (ABT)_{Both\text{-}tab\ cooling}$.

    Here ATD, RTD and ABT are axial temperature difference, radial temperature difference and average battery temperature, respectively.

b.  After 80 % DOD, the exponential rise in battery temperature is observed for all the geometries under all types of cooling approaches. In this regard, deep discharge of battery must be avoided to promote battery safety and lifecycle. For all the geometry, the temperature of battery cap and positive terminal is comparatively higher than that of other parts. From the analysis of volumetric heat generation rate, it has been observed that battery cap and Jelly-roll are primarily responsible for axial and radial temperature gradient respectively.

c.  Higher values of heat transfer coefficient reduce the battery temperature but increases the thermal heterogeneity inside the battery. Although, the effect of heat dissipation condition is different for different geometries, the percentage drop in temperature in case of forced cooling (for radial cooling) is maximum for LG battery i.e., 24.2 % for approach. This indicates that the effect (in terms of battery temperature) of increase in the heat transfer coefficient, is maximum for LG battery in case of radial cooling.

d.  In terms of temperature difference, while the values for radial temperature difference is found to be less for all the considered geometries under all the cooling approaches analysed, it is least for LG (for both-tab cooling). Hence, LG geometry is recommended for cooling systems with both-tab cooling approach. Considering the design criterion based on the average battery temperature, it is found to be least with mixed cooling approach and DG geometry is recommended for such cooling systems.

e.  The axial(radial) temperature difference is less under the radial(both-tab) cooling. This indicates that radial temperature uniformity is more in case of both-tab cooling than radial cooling, which is desired for higher life cycle of LIBs. However, both-tab cooling systems design has practical limitations due the presence of electrical connections in the tabs.

These important findings of the numerical analysis are anticipated as noteworthy aspects of battery geometry and its associated implication in the thermal management and battery life.

**Acknowledgement:**

A.M. would like to acknowledge the necessary software supports provided by Department of Chemical Engineering and Department of Applied Mechanics, Motilal Nehru National Institute of Technology Allahabad, India. S. G. want to acknowledge the European commission for the award of MarieSkłodowska-Curie Postdoctoral Fellowship (MSCA-PF: Grant number-ENHANCER-101067998). Lithium-ion battery as energy storage device is recognized as a part of clean technology but its increased usage without the availability of affordable, clean and renewable energy sources may not serve the purpose of its existence. Its associated adverse effect on the environment in terms of huge and toxic waste generation and increased carbon footprint needs immediate attention of energy policy makers. Till then, authors would also like to acknowledge those who try to minimize the usage of batteries for private vehicle to save our green earth. Authors would also like to dedicate this work to the inventor of Lithium-ion batteries, Professor John Bannister Goodenough, on his 100[th] birthday on 25[th] July 2022.

**Conflict of Interest:**

The authors declare no conflict of interest.

**Author Contributions:**

D.D.: contributed to the conceptualization, methodology, simulation, draft preparation, data curation; A.M.: contributed to the conceptualization, simulation, supervision, data curation, writing, reviewing and editing; S.G.: contributed to the drafting, writing, reviewing and editing; M.V.R.: Contributed to the writing, reviewing and editing, R.P.: Co-supervision, writing and editing. All authors reviewed before submission and approved for submission.

**Data Availability Statement**

The data that support the findings of this study are available from the corresponding author for open science.

**References**

[1] Reddy, Mogalahalli V., et al. "Brief history of early lithium-battery development." *Materials* 13.8 (2020): 1884.

[2] Goodenough, John B., and Kyu-Sung Park. "The Li-ion rechargeable battery: a perspective." *Journal of the American Chemical Society* 135.4 (2013): 1167-1176.

[3] Olivetti, Elsa A., et al. "Lithium-ion battery supply chain considerations: analysis of potential bottlenecks in critical metals." *Joule* 1.2 (2017): 229-243.

[4] Zubi, Ghassan, et al. "The lithium-ion battery: State of the art and future perspectives." *Renewable and Sustainable Energy Reviews* 89 (2018): 292-308.

[5] Ahmad, Taufeeq, et al. "Identifying Efficient Cooling Approach of Cylindrical Lithium-Ion Batteries." *Energy Technology* 10.2 (2022): 2100888.

[6] Goodenough, John B., and Youngsik Kim. "Challenges for rechargeable Li batteries." *Chemistry of materials* 22.3 (2010): 587-603.

[7] 7Doyle, Marc, Thomas F. Fuller, and John Newman. "Modeling of galvanostatic charge and discharge of the lithium/polymer/insertion cell." *Journal of the Electrochemical society* 140.6 (1993): 1526.

[8] Baird, Austin R., et al. "Explosion hazards from lithium-ion battery vent gas." *Journal of Power Sources* 446 (2020): 227257.

[9] Choudhari, V. G., et al. "Numerical investigation on thermal behaviour of $5 \times 5$ cell configured battery pack using phase change material and fin structure layout." *Journal of Energy Storage* 43 (2021): 103234.

[10] Kaliaperumal, Muthukrishnan, et al. "Cause and mitigation of lithium-ion battery failure—A review." *Materials* 14.19 (2021): 5676.

[11] Hou, Junxian, et al. "Thermal runaway of Lithium-ion batteries employing LiN (SO2F) 2-based concentrated electrolytes." *Nature communications* 11.1 (2020): 1-11.

[12] Ahmad, P., S. Shriram, and K. Gi-Heon. "Large Format Li-Ion Batteries for Vehicle." *Available: Applications (http://www. nrel. gov/docs/fy13osti/58145. pdf* (2013).

[13] Ma, Shuai, et al. "Temperature effect and thermal impact in lithium-ion batteries: A review." *Progress in Natural Science: Materials International* 28.6 (2018): 653-666.

[14] Forgez, Christophe, et al. "Thermal modeling of a cylindrical LiFePO4/graphite lithium-ion battery." *Journal of Power Sources* 195.9 (2010): 2961-2968.

[15] Wang, Helin, and Xueye Chen. "Numerical simulation of heat transfer and flow of Al2O3-water nanofluid in microchannel heat sink with cantor fractal structure based on genetic algorithm." *Analytica Chimica Acta* (2022): 339927.

[16] Hunt, Ian A., et al. "Surface cooling causes accelerated degradation compared to tab cooling for lithium-ion pouch cells." *Journal of The Electrochemical Society* 163.9 (2016): A1846.

[17] Ji, Yan, Yancheng Zhang, and Chao-Yang Wang. "Li-ion cell operation at low temperatures." *Journal of The Electrochemical Society* 160.4 (2013): A636.

[18] Lu, W., et al. "In Situ Thermal Study of Li1+ x [Ni1∕ 3Co1∕ 3Mn1∕ 3] 1− x O2 Using Isothermal Micro-clorimetric Techniques." *Journal of the Electrochemical Society* 153.11 (2006): A2147.

[19] Tranter, T. G., et al. "Probing heterogeneity in li-ion batteries with coupled multiscale models of electrochemistry and thermal transport using tomographic domains." *Journal of The Electrochemical Society* 167.11 (2020): 110538.

[20] Somasundaram, Karthik, Erik Birgersson, and Arun Sadashiv Mujumdar. "Thermal–electrochemical model for passive thermal management of a spiral-wound lithium-ion battery." *Journal of Power Sources* 203 (2012): 84-96.

[21] Al Hallaj, Said, and J. R. Selman. "A novel thermal management system for electric vehicle batteries using phase-change material." *Journal of the Electrochemical Society* 147.9 (2000): 3231.

[22] Tomaszewska, Anna, et al. "Lithium-ion battery fast charging: A review." *ETransportation* 1 (2019): 100011.

[23] Sun, Jieyu, et al. "Liquid cooling system optimization for a cell-to-pack battery module under fast charging." *International Journal of Energy Research* 46.9 (2022): 12241-12253.

[24] Ye, Yonghuang, et al. "Numerical analyses on optimizing a heat pipe thermal management system for lithium-ion batteries during fast charging." *Applied Thermal Engineering* 86 (2015): 281-291.

[25] Chen, Siqi, et al. "Multi-objective optimization design and experimental investigation for a parallel liquid cooling-based Lithium-ion battery module under fast charging." *Applied Thermal Engineering* 211 (2022): 118503.

[26] Chen, Siqi, et al. "A thermal design and experimental investigation for the fast charging process of a lithium-ion battery module with liquid cooling." *Journal of Electrochemical Energy Conversion and Storage*, *17*(2020).

[27] Sabbah, Rami, et al. "Active (air-cooled) vs. passive (phase change material) thermal management of high power lithium-ion packs: Limitation of temperature rise and uniformity of temperature distribution." *Journal of Power Sources* 182.2 (2008): 630-638.

[28] Duh, Yih-Shing, et al. "Comparative study on thermal runaway of commercial 14500, 18650 and 26650 LiFePO4 batteries used in electric vehicles." *Journal of Energy Storage* 31 (2020): 101580.

[29] Choudhari, V. G., et al. "Numerical investigation on thermal behaviour of 5× 5 cell configured battery pack using phase change material and fin structure layout." *Journal of Energy Storage* 43 (2021): 103234.

[30] Shukla, Vishesh, et al. "Size-dependent Failure Behavior of Lithium-Iron Phosphate Battery under Mechanical Abuse." *arXiv preprint arXiv:2206.11732* (2022).

[31] Abada, Sara, et al. "Safety focused modeling of lithium-ion batteries: A review." *Journal of Power Sources* 306 (2016): 178-192.

[32] Makinejad, Kamyar, et al. "A lumped electro-thermal model for Li-ion cells in electric vehicle application." *World Electric Vehicle Journal* 7.1 (2015): 1-13.

[33] Worwood, Daniel, et al. "A new approach to the internal thermal management of cylindrical battery cells for automotive applications." *Journal of Power Sources* 346 (2017): 151-166.

[34] Newman, John, and William Tiedemann. "Porous-electrode theory with battery applications." *AIChE Journal* 21.1 (1975): 25-41.

[35] Kala, Shashi, and A. Mishra. "Battery recycling opportunity and challenges in India." *Materials Today: Proceedings* 46 (2021): 1543-1556.

[36] S Kala, A Mishra, V Shukla, Indian Chemical Society,2020

[37] Li, Shen, et al. "Optimal cell tab design and cooling strategy for cylindrical lithium-ion batteries." *Journal of Power Sources* 492 (2021): 229594.

[38] Tran, Manh-Kien, et al. "Comparative study of equivalent circuit models performance in four common lithium-ion batteries: LFP, NMC, LMO, NCA." *Batteries* 7.3 (2021): 51.

[39] Chitta, SandeepDattu, et al. "Comparison of lumped and 1D electrochemical models for prismatic 20Ah LiFePO4 battery sandwiched between minichannel cold-plates." *Applied Thermal Engineering* 199 (2021): 117586.

[40] Yang, Xiaolong, et al. "Effect of ambient dissipation condition on thermal behavior of a lithium-ion battery using a 3D multi-partition model." *Applied Thermal Engineering* 178 (2020): 115634.

[41] Diaz, Laura Bravo, et al. "Measuring Irreversible Heat Generation in Lithium-Ion Batteries: An Experimental Methodology." *Journal of The Electrochemical Society* 169.3 (2022): 030523.

[42] Wang, Zhenpo, Jun Ma, and Lei Zhang. "Finite element thermal model and simulation for a cylindrical Li-ion battery." *IEEE Access* 5 (2017): 15372-15379.

[43] Panchal, S., et al. "Electrochemical thermal modeling and experimental measurements of 18650 cylindrical lithium-ion battery during discharge cycle for an EV." *Applied Thermal Engineering* 135 (2018): 123-132.

[44] Ye, Yonghuang, et al. "Effect of thermal contact resistances on fast charging of large format lithium ion batteries." *Electrochimica Acta* 134 (2014): 327-337.

[45] Xie, Yi, et al. "An improved resistance-based thermal model for a pouch lithium-ion battery considering heat generation of posts." *Applied Thermal Engineering* 164 (2020): 114455.

[46] Mohammadian, Shahabeddin K., and Yuwen Zhang. "Thermal management optimization of an air-cooled Li-ion battery module using pin-fin heat sinks for hybrid electric vehicles." *Journal of Power Sources* 273 (2015): 431-439.

[47] Zohuri, Bahman. "Forced convection heat transfer." *Thermal-Hydraulic Analysis of Nuclear Reactors*. Springer, Cham, 2017. 323-345.

[48] Zhang, Guangsheng, et al. "In situ measurement of radial temperature distributions in cylindrical Li-ion cells." *Journal of The Electrochemical Society* 161.10 (2014): A1499.

[49] Wu, Billy, et al. "Coupled thermal–electrochemical modelling of uneven heat generation in lithium-ion battery packs." *Journal of Power Sources* 243 (2013): 544-554.